\documentclass[%
 reprint,
nofootinbib,
 amsmath,amssymb,
 aps,
]{revtex4-1}

\usepackage{graphicx}
\usepackage{dcolumn}
\usepackage{bm}

\begin{document}


\title{Regularised Kalb-Ramond Magnetic Monopole with Finite Energy}

\author{Nick E Mavromatos}
\affiliation{%
Theoretical Particle Physics and Cosmology Group, Department of Physics, King's College London, Strand, London WC2R 2LS, UK
}%

\author{Sarben Sarkar}
\affiliation{%
Theoretical Particle Physics and Cosmology Group, Department of Physics, King's College London, Strand, London WC2R 2LS, UK
}%

\begin{abstract}
In a previous work we suggested a self-gravitating electromagnetic monopole solution in a string-inspired model involving global spontaneous breaking of a $SO(3)$ internal symmetry and Kalb-Ramond (KR) axions, stemming from an antisymmetric tensor field in the massless string multiplet. These axions carry a charge, which, in our model, also plays the r\^ole of the magnetic charge. The resulting geometry is close to that of a Reissner-Nordstr\"om  (RN) black hole with charge proportional to the KR-axion charge. We proposed the existence of a thin shell structure surrounding a (large inner) core as the dominant mass contribution to the energy functional. Although the resulting energy was finite, and proportional to the KR-axion charge, the size of the shell was not determined and left as a phenomenological parameter. In the current article, we propose a new way to calculate the size of the thin shell: string theory considerations suggest that the short-distance physics inside the inner core may be dominated by a positive cosmological constant term proportional to the scale of the spontaneous symmetry breaking of $SO(3)$.The size of the shell is estimated by matching the RN metric of the shell to the de Sitter metric inside the core. The matching entails the Israel junction conditions for the metric and its first derivatives at the inner boundary of the shell, and determines the inner mass-shell radius. The axion charge plays an important r\^ole in guaranteeing the positivity of the ``mass coefficient'' of the gravitational potential term appearing in the metric; so, the KR electromagnetic monopole shows normal attractive gravitational effects. This is to be contrasted with the axion-less global monopole case
 where such a matching is known to yield a negative ``mass coefficient'' (and, hence, a repulsive gravitational effect). The total energy of our monopole within the shell is calculated.  As a result of the violation of Birkhoff's theorem (due to the formal divergence of the energy functional in the absence of a large distance cut-off), the total energy does not have to equal the mass coefficient. However, for phenomenologically relevant sets of parameters, the ratio of the total energy and the mass coefficient in the shell is close to 1. The  gravitational ``effective mass coefficient''  in the shell can be made equal to the total energy outside the core by a small decrease in the cosmological constant in the de Sitter region. This is achieved through a dilaton potential which is suitably negative inside the de-Sitter region, but vanishes outside that region.

\end{abstract}

\maketitle


\section{\label{sec:level1} Introduction: A review of the Model }

Recently~\cite{msmono} the authors have formulated a novel magnetic monopole solution in a string-inspired model, involving the coupling of self-interacting scalar fields, responsible for the spontaneous breaking of a global SO(3) symmetry, with a Kalb-Ramond (KR) field, associated with the spin-one antisymmetric tensor field of the string gravitational multiplet. The model generalises the gravitational monopole model \cite{vilenkin} and contains, in addition, electromagnetic $U(1)$ gauge fields, which couple to the KR field via a dilaton field- the dilaton is the scalar field of the string gravitational multiplet.

The  low energy effective Lagrangian \footnote{The Lagrangian is inspired by perturbative weakly coupled string theory.} of the model is 
 \begin{equation}
\label{Lagrangian}
L = {\left( { - g} \right)^{1/2}}\left\{ \begin{array}{l}
\frac{1}{2}{\partial _\mu }{\chi ^A}{\partial ^\mu }{\chi ^A} - \frac{\lambda }{4}{\left( {{\chi ^A}{\chi ^A} - {\eta ^2}} \right)^2} - R\\
 + \frac{1}{2}{\partial _\mu }\Phi {\partial ^\mu }\Phi  - V\left( \Phi  \right)\\
 - \frac{1}{{12}}{e^{ - 2\Phi }}{H_{\rho \mu \nu }}{H^{\rho \mu \nu }} - \frac{1}{4}{e^{ - \Phi }}{f_{\mu \nu }}{f^{\mu \nu }}
\end{array} \right\}.
\end{equation}
In the above expression, $-g$ denotes the determinant of the metric, $\chi^A$; $A=1,2,3,$ are the scalar fields responsible for the spontaneous breaking of the SO(3) internal symmetry via their vacuum expectation value (v.e.v.) $\eta$; $H_{\mu\nu\rho}=\partial_{[\mu} \, B_{\nu\rho]}$ is the totally antisymmetric field strength of the antisymmetric tensor field $B_{\mu\nu}=-B_{\nu\mu}$ (with $[\dots]$ denoting antisymmetrisation of the respective indices); $f_{\mu\nu}$ is the Maxwell tensor of the electromagnetic $U(1)$ gauge field;
$\Phi$, the dilaton field, is stabilised (in perturbative string theory) to a constant value $\Phi=\Phi_0$ at the minimum of $V(\Phi)$ with $V(\Phi_0)=0$.
 For a core region surrounding the monopole centre where gravity is strong, the underlying string theory may be strongly coupled. In this core region the behaviour of the dilaton may result in a regularisation of the associated space-time singularity~\cite{msmono} of the monopole. This regularisation and its ramification is the subject of this paper. 

For the lowest order string effective action~\cite{string,zwi} in terms of the gravitational multiplet, it is known that the field strength $H_{\mu\nu\rho}$ plays the r\^ole of a totally antisymmetric torsion in a generalised Christoffel connection. In \emph{four} space-time dimensions, the dual of the torsion is a pseudoscalar (KR axion) field $b(x)$~\cite{aben},
\begin{equation}\label{dH}
H_{\mu\nu\rho} = \epsilon_{\mu\nu\rho\sigma}  \, e^{2\Phi} \, \partial^\sigma \, b(x)
\end{equation}
with $ \epsilon_{\mu\nu\rho\sigma}$ the covariant Levi-Civita antisymmetric symbol. As discussed in \cite{msmono}, the classical radial solution for $b(r)$ reads~\footnote{In \cite{msmono} we fixed the dilaton to a constant background $\Phi_0$ such that the string coupling $g_s=e^{\Phi_0}=\frac{1}{\sqrt{2}}$, which is compatible with the order of magnitude of string couplings $\frac{g_s^2}{4\pi} =\frac{1}{20}$ characteristic of phenomenologically relevant models, . We also absorbed the factor $e^{-\Phi_0/2}$ in a redefinition of the electromagnetic field strength $f_{\mu\nu}$, so that one has a canonically normalised Maxwell term  in (\ref{Lagrangian}). It is with these conventions that the normalization of the axion charge $\zeta$ in (\ref{bfield}) is fixed. An arbitrary constant value of $\Phi_0$ can thus be absorbed in an appropriate normalization of $\zeta$.}:
\begin{equation}
b'\left(r\right)=\frac{\zeta}{r^{2}}\sqrt{\frac{A\left(r\right)}{B\left(r\right)}}\label{bfield}
\end{equation}
where $r$ is the radial distance (with dimension of length) from the 
centre of the configuration. The quantity 
$\zeta$ is a \emph{dimensionless} constant of integration which measures the strength
of the KR field strength and the pseudoscalar field $b(x)$  has mass dimension one.  We call $\zeta$ the KR-axion charge. 

The (dimensionless) quantities
$A(r) $ and $\, B(r)$ are radial functions appearing in the solution for the metric tensor~\cite{msmono} 
\begin{equation}
g_{\mu\nu}=\left(\begin{array}{cccc}
B(r)\\
 & -A(r)\\
 &  & -r^{2}\\
 &  &  & -r^{2}\sin^{2}\theta
\end{array}\right)~.\label{eq:19}
\end{equation}
We have from~\cite{msmono}: $A(r) \, B(r) = 1 + {\mathcal O}(r^2)$, for $r \to 0$, while $A(r) \, B(r) = 1 + {\mathcal O}(r^{-2})$ for $r \to \infty$. 
In the current work we shall adopt the following approximate relations for the entire range of $r$
\begin{eqnarray}\label{ab1}
A\left( r \right)B\left( r \right) &\approx& 1~, \nonumber \\
B\left( r \right) &=& 1 - 8\pi {\rm G}\, \eta^2 - \frac{{2m {\rm G}}}{r} + \frac{8\pi {\rm G}\, p}{{{r^2}}} ~,
\end{eqnarray}
where $m$ is the Schwarzschild mass of the monopole, $r$ is the radial distance from the centre, both having the appropriate dimensions and G is Newton's gravitational constant of four dimensional space time. 
For the solution~\cite{msmono} of phenomenological relevance to current colliders (i.e. with a detectable monopole mass): if $8\pi {\rm G}\, \eta^2 \ll 1$, with $\eta $ assumed  to be much lower than the Planck scale, the monopole might have a mass ${\mathcal O}(10\, {\rm TeV}) $.
The solution (\ref{ab1}) is the Reissiner-Nordstr\"om (RN) expression~\cite{RNsol} for a magnetic black hole and is compatible with the asymptotic forms studied in \cite{msmono} and so the expression (\ref{ab1}) provides a good approximation for our purposes. From (\ref{bfield}) and (\ref{ab1})  we obtain the following solution for the KR axion
\begin{equation}
\label{bprime}
b'\left( r \right) = \frac{\zeta }{{{r^2}B\left( r \right)}}.
\end{equation}
It will be convenient to use dimensionless variables, and so we will now work in units of $8\pi  {\rm G}=1$. 
In these units the metric function  $B(r)$ (\ref{ab1}) become:
\begin{eqnarray}\label{metric_units}
B\left( r \right) &=& 1 -  \eta^2 - \frac{2M}{r} + \frac{p}{r^2} \nonumber \\
 p &=& 2\zeta^2 ~, \quad M \equiv \frac{m}{8\pi}, \quad r \rightarrow  \frac{r}{\sqrt{8\pi}}~,
\end{eqnarray}
with $r $, $M$ and $\eta$ dimensionless (or, equivalently, expressed in reduced Planck mass scale units, in which the Planck mass is $M_P= \ell_P^{-1} = \sqrt{8\pi}$, with $\ell_P$ the Planck length).

As discussed in detail in \cite{msmono}, the solution for the 
electromagnetic $U(1)$ Maxwell tensor is~\cite{msmono}:
\begin{equation}
f_{\mu\nu}=\left(\begin{array}{cccc}
0 & 0 & 0 & 0\\
0 & 0 & 0 & 0\\
0 & 0 & 0 & 2r\sin\theta\, W(r)\\
0 & 0 & -2r\sin\theta\, W(r) & 0
\end{array}\right)~,
\label{eq:20}
\end{equation}
with 
\begin{equation}\label{wform}
W\left( r \right) = \frac{\zeta}{r}~,
\end{equation}
for all $r$. 

The associated magnetic field has only a radial component, which in \emph{contravariant} form reads~\cite{msmono}:
\begin{equation}\label{magnetic}
\mathcal{B}^r = \frac{1}{\sqrt{-g}} \, \eta^{r \theta \phi}\, f_{\theta\phi} = \sqrt{\frac{2}{A\, B}} \, \frac{W(r)}{r} \simeq \frac{\sqrt{2}\, \zeta}{r^2}~,
\end{equation}
where we took into account Eq.~(\ref{eq:19}) and (\ref{ab1}).  From this it follows that the magnetic charge is $g= \pm \sqrt{2}\, \varsigma$. The electric field and charge are zero. 
It is evident from (\ref{ab1}) that the constant $p$  is the square of the magnetic charge, 
\begin{equation}\label{pdef}
p= g^2 = 2\, \zeta^2,
\end{equation}
and thus the KR-axion charge provides the (Dirac) magnetic charge in this model~\cite{msmono}, via the RN geometry (\ref{ab1}) ((\ref{metric_units})),
\begin{equation}\label{magch}
g=\sqrt{2}\, \zeta~.
\end{equation}
Dirac quantization then leads to large values of the KR axion charge $\zeta$ (in our natural units), since the Dirac quantisation condition is 
\begin{equation}\label{dirac}
g \, e = \frac{n}{2}, \quad n \in {\tt Z}^{+} ~.
\end{equation}
The global gravitational monopole is known to modify the four-dimensional asymptotic space-time  to that of a curved space-time with a conical singularity, corresponding to a deficit angle $8\pi \, {\rm G} \, \eta^2 $ (\emph{cf.} (\ref{ab1})
with scalar curvature  
\begin{equation}\label{curvsing}
R \propto \frac{16\pi \, {\rm G} \, \eta^2}{r^2}~. 
\end{equation}
Similar features remain for the RN geometry in \cite{msmono}.

For the scalar triplet field, associated with the spontaneous symmetry breaking of the global SO(3) symmetry, we made the ansatz~\cite{msmono}
\begin{equation}\label{scalar}
{\chi ^A}\left( r \right) = \eta f\left( r \right)\frac{{{x^A}}}{r}\,, \quad A=1,2,3~,
\end{equation}
where $x^A$, $A=1,2,3$ are Cartesian spatial coordinates, with the asymptotic behaviour 
\begin{eqnarray}\label{asympt}
f(r \to 0) &\simeq &  f_0 \, r \to 0 \, \quad f_0 ={\rm constant} \in {\tt R}~, \nonumber \\
f(r \to \infty )  &\to&  1~. 
\end{eqnarray}

In \cite{msmono} we adopted a heuristic approach to demonstrating the finiteness of the monopole mass, based on the assumption of a bag-like structure. Specifically, 
we assumed that the entirety of the mass resides inside a bag of radius $L_c$, and, in fact, within a thin shell bounded by the (large) shell radius, yielding the following estimate for the total energy of the magnetic monopole~\cite{msmono}:
\begin{eqnarray}
\label{energy}
{\mathcal E} 
\simeq  4\pi \int_{\alpha\, L_c}^{L_c}  {dr\;{r^2} \, 
\left[ \begin{array}{l}
\frac{{2W{{\left( r \right)}^2}}}{{B\left( r \right){r^2}}} + \frac{{b'{{\left( r \right)}^2}}}{{4}} + \\
\eta^2\, \left( {\frac{{f{{\left( r \right)}^2}}}{{B\left( r \right){r^2}}} + \frac{{f'{{\left( r \right)}^2}}}{{2}}} \right) + \\
\frac{{{\lambda \, \eta^4}}}{{4 B\left( r \right)}}{\left( {f{{\left( r \right)}^2}-1} \right)^2} 
\end{array} \right]} , 
\end{eqnarray}
where $0 < \alpha < 1$ is a dimensionless phenomenological  parameter determining the mass shell thickness. Inside the shell, there are non-trivial configurations of the KR axion, electromagnetic field, and the scalar triplet fields $\chi^A, \, A=1,2,3$ responsible for the SO(3) spontaneous symmetry breaking in the model. 
Upon the assumption of large $L_c $ and $\alpha \, L_c$ (as compared to the Planck length), one obtains from (\ref{energy})
\begin{eqnarray}\label{energy2}
{\mathcal E} &\simeq & \frac{1}{\alpha}\, (1 - \alpha) \, \Big(9\pi \, \zeta^2 - \frac{4\pi}{\lambda}\Big)\, \frac{1}{L_c} + 4\pi \, \eta^2 \, (1 - \alpha)\, L_c ~,  \nonumber \\
&& L_c, \, \alpha L_c \gg 1 ~.
\end{eqnarray}
In arriving at the above result we used (\ref{wform}), as well as the asymptotic behaviour of the scalar triplet $\chi^A, \, A=1,2,3$, for $\quad \lambda \, \eta^2 \, r \gg 1$, \emph{i.e.} $f(r) \simeq 1 - \frac{2}{\lambda \, \eta^2\, r^2 }$, found in~\cite{msmono}. As in \cite{msmono} we will ignore the $1/\lambda$ terms in (\ref{energy2}) since we are
working in the large $\lambda \gg 1$ limit; we note the relative minus sign of this term relative to the $\zeta^2$ term in (\ref{energy2}). (This observation also corrects a typographical error in the corresponding formula for the total monopole energy in \cite{msmono}.) 

We have assumed (as in \cite{msmono}) that the solution (\ref{wform}) for the electromagnetic field is valid for both small and large $r$. This follows
from the dilaton equation of motion of the original Lagrangian (\ref{Lagrangian}) of the model~\cite{msmono}, which requires that the (covariant) square of the electromagnetic tensor $f_{\mu\nu} f^{\mu\nu}$ is proportional
to the KR kinetic term $\partial_\mu b\, \partial^\mu b$. Since the solution for the $b$-field (\ref{bfield}) is valid uniformly for $r$ (both small and large $r$ , compared to the Planck length), the kinetic term for $b$ is $\zeta^2/r^4$ in leading order for the two limits; the ansatz (\ref{eq:20}) then implies the validity of (\ref{wform}) for both large and small $r$. It should also be remarked that the metric in \cite{msmono} is not rigorously RN, since the product $A(r) \, B(r)$ is not exactly 1 for all $r$, but resembles a RN space-time to a very good approximation in the limit of both large and small $r$. Moreover, these two  asymptotic RN space-times are characterised by charge parameters, which differ from each other by terms 
of order $1/\lambda$; thus only in the limit $\lambda \to \infty$ does one obtain the same RN metric in the $r \to 0$ and $r \to \infty$ limits~\cite{msmono}.
In view of the small-$r$ regularization of the KR self-gravitating monopole solution (using de Sitter space-time in a core region)  to be discussed  in the next section \ref{sec:reg},  we will only be interested in the large $r$ limit. The radius $\delta$ of the core region will turn out to be  much larger than the Planck length scale but, in a string theory with large extra dimensions, it can be of order of the string scale~\cite{string, zwi}. 

The assumption of a single RN space-time outside the de Sitter region is valid for all $\lambda$. We assumed in \cite{msmono} strong coupling $\lambda \gg 1$ in order to ensure that the scalar triplet fields are near their vacuum expectation value, i.e. large quantum fluctuations are suppressed in the respective path integral; we obtain $L_c$ by minimizing the right-hand side of (\ref{energy2}) with respect to $L_c$:
\begin{equation}\label{corR}
L_c \simeq \frac{3}{2} \sqrt{\frac{1}{\alpha}} \, \frac{|\zeta|}{\eta} ~,
\end{equation}
which implies 
\begin{equation}\label{monoenergy}
{\mathcal E} \simeq 12\pi \, \sqrt{\frac{1}{\alpha}} \, (1 - \alpha) \, |\zeta|\, \eta ~.
\end{equation}
In \cite{msmono} the parameter $\alpha$ could not be estimated. In general, $\sqrt{\alpha}$, and hence the size of the shell, could depend on the
coupling $\lambda$, but it was the assumption in \cite{msmono} that any potential $\lambda$-dependence in $\alpha$ was such that the terms of order $\mathcal{O}(\frac{1}{\lambda})$ (that were ignored in deriving (\ref{corR}) and (\ref{monoenergy})) were subleading.    

 It is the purpose of this note to attempt a resolution of these important issues by demonstrating in detail the emergence of such bag-like shell structures 
 in our self-gravitating monopole solution on implementing a string theory inspired \emph{regularization} of the physical singularity at $r\to 0$ (\ref{curvsing}). This regularisation allows us to 
 calculate the finite monopole mass in terms of the parameters of the model (\ref{Lagrangian}), by providing a natural estimate of  the parameter $\alpha$.

\section{Regularising the Curvature Singularity of the KR electromagnetic monopole \label{sec:reg}}

\begin{figure}[h!]
\begin{center}
\vspace{0.5cm}
\includegraphics[scale=0.8]{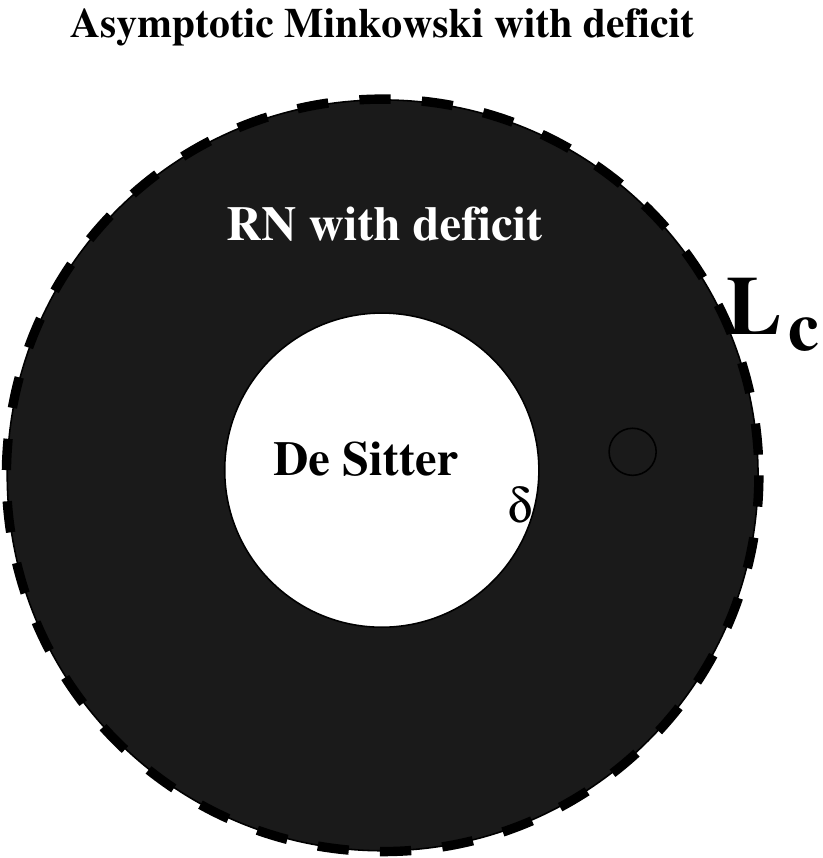}
\end{center}
\caption{The regularised seff-gravitating KR electromagnetic monopole. Israel matching conditions apply at the inner shell radius $\delta$ between the inner-shell de Sitter space time and 
the exterior RN space time with an angular deficit proportional to the SO(3) spontaneous symmetry breaking scale $\eta^2$. The dark shaded area depicts the area where most of the 
mass of the monopole lies. The outer radius of the shaded boundary,  $L_c$, is sufficiently large (compared to the Planck length), so that, at  this boundary, the RN space time with the angular deficit $\propto \eta^2$ is, to a very good approximation, an asymptotic Minkowski space time with the same angular deficit. Hence no additional matching is required at the core radius $L_c$, as the RN space extends formally from $\delta$ to infinity. This is indicated by the dashed-line outer boundary (at $L_c$) of the dark shaded region.}\label{RNshell}
\end{figure}

For a self-gravitating KR electromagnetic monopole~\cite{msmono}, the metric assumes the static RN form (\ref{eq:19}), (\ref{ab1}), with an angular deficit proportional to $\eta^2$. 
We may regularise the associated curvature singularity (\ref{curvsing}) by considering a (small) region around the singularity and replacing the space-time inside it by an appropriate space-time
of de Sitter type of radius $\delta$. (See Fig.~\ref{RNshell}, where the white (dark) shaded region corresponds to the de Sitter (RN with angular deficit) space-times.). The outer space-time corresponds to the RN metric (\ref{eq:19})~\footnote{In the context of our string/brane-theory inspired model, with its low-energy Lagrangian (\ref{Lagrangian}), this regularisation may be understood as follows: the regularisation concerns the ultra-violet region near the centre of the black hole/monopole, $r \to 0$, where gravity is strong and so a strongly coupled string theory might be expected, with a string coupling $g_s (r \to 0) = e^{\Phi(r \to 0)} \gg  1$. Consequently, a non-constant nature of the dilaton, ignored in our analysis outside the core region, becomes important.  From the effective Lagrangian (\ref{Lagrangian}) we note that the dilaton does \emph{not} couple directly to the terms involving the scalar triplet fields $\chi^a$, $a=1,2,3$. 
As a result, although the scalar fields go to zero as $r \to 0$, the contribution of the term $\frac{1}{4}\lambda \, \eta^4$  from the scalar potential remains, and is independent of the dilaton.  The antisymmetric tensor and electromagnetic terms, on the other hand, are suppressed due to the dilaton prefactors (which are inversely proportional to powers of the string coupling). This is implied by (\ref{Lagrangian}) on using the dilaton equation of motion and 
requiring that, in the strong coupling limit of string theory, the field strength $H_{\mu\nu\rho}$ (\ref{dH}) remains finite. Furthermore, in the de Sitter region, higher-order derivative terms in the string effective action become important, whose form leads for  to Born-Infeld type electromagnetism~\cite{string}. Both the dilaton prefactors and the Born-Infeld form of the electromagnetic terms imply a subdominant r\^ole compared to that of the $\frac{1}{4} \, \lambda \, \eta^4$ in the effective action
(see ref.~\cite{zwi} for more details). One might expect that strong string couplings characterise length scales of order of the string length, and indeed the radius of the de Sitter-domination region $\delta$ may be identified with the corresponding string length scale $\ell_s =1/M_s$, where $M_s$ is the corresponding string mass scale. As we shall see later on, the scale $\delta$ is much larger than the four-dimensional Planck length, which implies a large string length scale. 
However, the above arguments are rather heuristic and 
valid only within the context of our low energy effective Lagrangian (\ref{Lagrangian}), which describes the \emph{tree-level} dynamics of an underlying string theory model. Hence they do not constitute a ``derivation'' of our regularisation procedure in the context of a microscopic string theory model. For strong string couplings, \emph{string loop corrections} become important, and the non-perturbative action is not known in closed form. However, for our purposes of regularising the black hole singularity, such heuristic constructions suffice to give a plausible justification for using de Sitter space as a way of regularising the black-hole singularity in our string-inspired field theory model.}. The cut-off radius $\delta$ of the appropriate boundary that separates the two space times may be determined by employing the well-known procedure initiated by Israel~\cite{israel} based on the matching of two spherical regions in space, an inner and an outer one, described by different metrics. The regions are separated by a ``thin-shell". We assume, following \cite{zaslas}, that the energy-momentum tensor vanishes on this interface hyperplane, and so no energy flow occurs through the boundary surface at $r=\delta$.

The matching conditions (``Israel conditions'') amount to demanding the 
continuity of the metric and its derivatives on the thin shell. This implies continuity of the curvature and thus (in view of the Einstein equations) the stress-energy tensor of the model.
This procedure was applied to a regularisation of the self-gravitating global monopole of \cite{vilenkin} in \cite{lousto}, and for the conventional RN metric in \cite{zaslas} and \cite{rnreg}. (In  \cite{rnreg} it was also argued that, for (restricted) stability, the perturbed shell has to satisfy a certain polytropic equation of state). For a different approach to the regularisation of black-hole singularities see \cite{klink}. 

\subsection{Israel Conditions for Regularised KR Monopole}

In our bag-like model for the KR magnetic monopole~\cite{msmono}, the outer core radius $L_c$ is determined by minimisation of the energy functional integrated over the shell region between the radius $\delta$ (to be identified with the lower integration limit $\alpha L_c$) and $L_c$ in (\ref{energy}). The radius $L_c$ is assumed to be sufficiently large to act as an upper spatial cut-off for the otherwise divergent energy of the monopole. This divergence (which is shared with the global monopole case of \cite{vilenkin}) is due to the existence of the deficit $\eta^2$ in the induced space-time (which differentiates it from the standard Minkowski space-time). 
No matching  conditions are required on the outer core surface of radius $L_c \gg 1$, since the RN space-time with deficit extends formally to infinity.

In terms of spherical polar coordinates $r$, $\theta$, $\varphi$ we consider the following infinitesimal line element for the regularised space-time:
\begin{equation}\label{lineele}
ds^2_{\rm reg} = f(r) \,  dt^2  - \frac{dr^2}{f(r)} - r^2 (d\theta^2 + {\rm sin}^2\theta \, d\varphi^2),
\end{equation}
where the distribution function $f(r;\delta)$ is defined as follows:
\begin{eqnarray}\label{distr}
f(r; \delta) &=& B_1(r)\, \Theta( \delta - r) + B(r) \, \Theta (r - \delta), \nonumber \\
B_1(r) &=& 1 - \frac{1}{3}\Lambda\, r^2, \quad \Lambda > 0,
\end{eqnarray}
with $B(r)$ given in (\ref{metric_units}) and the positive parameter $\Lambda$ will be chosen later. 
In our static case,
the Israel matching conditions are:
\begin{eqnarray}\label{Israel1}
B_1(r=\delta) &=& B(r=\delta)~, \nonumber \\
\frac{d}{dr} B_1(r) \Big|_{r=\delta} &=& \frac{d}{dr} B(r) \Big |_{r=\delta}~.
\end{eqnarray}
which imply
\begin{eqnarray}\label{Israel2}
\frac{\Lambda}{3} &=& \eta^2 + \frac{2\, M}{\delta} - \frac{2\,\zeta^2}{\delta^2}~,  \quad  \nonumber \\
-\frac{2}{3}\, \Lambda \, \delta &=&   \frac{2\, M}{\delta^2} - \frac{4\,\zeta^2}{\delta^3}~.
\end{eqnarray}
These determine the cut-off radius $\delta >0$ and the ``mass coefficient'' $M$ of the metric (\ref{metric_units}) in terms of the de Sitter parameter $\Lambda > 0$~\footnote{As can be readily seen from (\ref{Israel2}), the interior region of radius $\delta$ (Fig.~\ref{RNshell}) cannot be a flat empty space with zero cosmological constant
.}:
\begin{eqnarray}\label{delM}
\delta &=& \frac{\eta}{\sqrt{2 \, \Lambda}}\, \Big(1 + \sqrt{1 + \frac{8\, \zeta^2\, \Lambda}{\eta^4}}\Big)^{1/2} > 0, \nonumber \\
2M &=& - \frac{2}{3}\, \Lambda \, \delta^3 + \frac{4\, \zeta^2}{\delta}~.
\end{eqnarray}
We note from (\ref{delM}) that in the \emph{absence} of the KR-axion charge (i.e. $\zeta=0$,  the global monopole case~\cite{vilenkin}), the mass coefficient $M = -\frac{2\Lambda}{3}\delta^3 < 0$ is \emph{negative}; since this coefficient appears in the gravitational potential, this case implies repulsive gravitational effects of the global monopole, as discussed in~\cite{lousto}. The situation changes drastically for the KR electromagnetic monopole case~\cite{msmono}; since, as can be deduced from (\ref{delM}), in the presence of sufficiently large $\zeta^2$, (necessary for 
the Dirac quantization condition (\ref{dirac})), strong coupling and appropriate values of $\Lambda > 0$ and $\delta >0$~\footnote{$\delta$ is not necessarily small, as we shall see below.}  it is quite possible that the negative term in the right-hand side of (\ref{delM}) is subdominant when compared to the positive term. In this case one could  obtain $M >0$, thus implying normal attractive gravitational effects for our KR monopole. This is indeed the case in our problem, as we show in the next subsection. 

Before doing so, it must be noted that the presence of a 
de Sitter (positive cosmological constant) space-time in the interior region $0 < r < \delta$ implies \emph{stability} of the resulting KR monopole, as a consequence of a \emph{balance} between the 
de-Sitter-induced \emph{repulsive} forces and the positive-mass-induced (normal gravitational) \emph{attractive} forces on the surface $\delta$. This contrasts with the situation for the the original global monopole solution of \cite{vilenkin} where there is still an ongoing debate~\cite{debate} on its stability. We may therefore consider the de Sitter regularization as a necessary physical property of our self-gravitating KR monopole which can guarantee its stability. 

\subsection{Regularisation Scheme : Similar to Global Monopole case}

To this end, we first require that the regularising cosmological constant $\Lambda$ in the inner de Sitter space-time coincides with the vacuum (dark) energy of the gravitational Lagrangian (\ref{Lagrangian}) in the absence of any other matter fields apart from the scalars $\chi^A$ (that is, ignoring KR axions and electromagnetic fields, for reasons stated previously). A vacuum energy $\Lambda$ arises on noting that that $\chi^A \to 0$ in the inner de Sitter region, compatible with the small $r$ behaviour (\ref{asympt}). This interpretation ensures the same regularization process in the cases of both global and KR electromagnetic monopoles.
We can then identify~\cite{lousto}:
 \begin{equation}\label{Lident}
 \Lambda = \frac{1}{4} \lambda \, \eta^4 > 0, \quad {\rm for} \quad  \lambda > 0~.
 \end{equation}
The following regime of parameters characterises the phenomenologically interesting KR electromagnetic monopole of \cite{msmono}, which we now use:
\begin{equation}\label{param}
\lambda \gg 1, \quad |\zeta| \gg 1, \quad \lambda \, \zeta^2 \gg 1, \quad \eta^2 \ll 1, \quad \lambda \eta^2 \ll 1~.
\end{equation}
In this regime, we obtain from (\ref{delM}) and (\ref{Lident}):
\begin{eqnarray}\label{delMfinal}
\delta & \sim & 2^{3/4} \, |\zeta |^{1/2} \, \lambda^{-1/4} \, \eta^{-1} \nonumber \\
&=&  2^{3/4} \, \frac{|\zeta |}{\eta} \, \big(\zeta^2 \, \lambda \big)^{-1/4} \,  \gg 1~, \nonumber \\
0 < M = \frac{m}{8\pi} &\sim & 0.79 \, |\zeta| \, \eta  \big(\lambda \, \zeta^2 \big)^{1/4}~, 
\end{eqnarray}
where we took into account (\ref{param}) and (\ref{metric_units}). 
We thus observe that in this case, both the cut-off $\delta$ and the mass coefficient $M$ are proportional to the KR-axion charge $\zeta$ (or equivalently the magnetic charge of the monopole), and $M $ is positive, implying normal (\emph{attractive}) gravitational effects, in contrast to the regularised global monopole case~\cite{lousto}. Phenomenologically we are interested in 
$M \ll 1$, which can be arranged for sufficiently small $ 0 < \eta  \ll 1$. Taking into account (\ref{param}), we observe that the monopole mass is much larger than the SO(3) spontaneous symmetry breaking scale, $M \gg \eta$. 

By identifying 
\begin{equation}\label{dident}
\delta = \alpha \, L_c
\end{equation}
 we can obtain from (\ref{delMfinal}) and (\ref{corR}) the following estimate for the parameter $0 < \alpha < 1$:
\begin{equation}\label{alfa}
0 < \alpha = 1.26 \, \big(\lambda \, \zeta^2 \big)^{-1/2}  \ll 1~,
\end{equation}
in view of (\ref{param}). 

Despite the smallness of $\sqrt{\alpha}$, however, both the cutoff $\delta =\alpha\, L_c $ and the shell radius $L_c$ are much larger than the Planck scale, consistent with the assumptions and estimates of \cite{msmono}. Moreover, since $\sqrt{\alpha } \propto \lambda^{-1/4}$, ignoring terms of ${\mathcal O}(\frac{1}{\lambda})$ for strong coupling $\lambda$ in the estimate (\ref{corR}) is consistent. Thus, although the bag-like  RN shell (dark shaded region in Fig.~\ref{RNshell}) is not so thin, nonetheless the most significant contributions to the energy integral (\ref{energy}) come indeed from large radial distances in the integrand, in qualitative agreement  to the estimates in \cite{msmono}.

With the value of $\alpha \ll 1 $ (\ref{alfa}), the total energy (\ref{energy2}) becomes to leading order:
\begin{equation}\label{energy3}
{\mathcal E} \simeq  8\pi \, (1.34 \, \big(\lambda \, \zeta^2 \big)^{1/4} |\zeta | \, \eta )~,
\end{equation}
where we pulled out explicitly the coefficient $8\pi$, which facilitates  a direct comparison with the monopole mass $m = 8\pi \, M$ (\ref{delMfinal}). From (\ref{energy3}), (\ref{delMfinal}) we obtain for 
the ratio  
\begin{equation}\label{ratio}
\mathcal E/ m \sim 1.7 ~,
\end{equation}
implying that, as a result of the divergent nature of the energy due to the angular deficit $\eta^2$ in the asymptotic space-time, the mass coefficient appearing in the gravitational potential is different from the total energy (which in a flat space time would be considered as the total monopole rest mass).  This violation of the weak equivalence principle and the invalidity of Birkhoff's theorem (due to the linear dependence of $\mathcal E$ on the cut-off $L_c$) are related. However, the order of magnitude of both terms in the regularised black hole is the same; this is to be expected  for large cores, since the space-time in their exterior is practically flat Minkowski (with a small deficit angle, see Fig.~\ref{RNshell}).

We should remark at this point that  for an extended object, such as the regularised monopole, we should be careful to use the `correct type' of mass in the presence of a gravitational field. In the next section we shall elaborate on this issue by defining properly the concept of an `effective mass' for our monopole solution in the presence of the de-Sitter regularising core region.

\section{Effective Mass concept in the regularised KR Monopole space-time} 

We would like to place the above results, especially (\ref{ratio}),  within the framework of standard concepts of mass in general relativity. In \cite{marsh} it was pointed out that in the (conventional) RN solution to the Einstein field equations, charge, like mass, admits a space-time signature, since it induces curvature of space-time. 
In view of the proportionality of the charge $\zeta$ to the total mass $M$ (\ref{delMfinal}) or energy $\mathcal E$ (\ref{energy3}), this is exactly what happens in our 
self-gravitating KR magnetic monopole case. 

In~\cite{marsh} it is argued that a spatial spherical surface of radius $R$, with the singularity of the RN located at the centre of the sphere, and argues that one can define two kinds of ``effective mass'' consistent with general relativity: one is an effective mass $M_{\rm eff}^{\rm Int}$ enclosed by the surface of radius $R$ (that is, associated with the inner region corresponding to radial distances $r < R$) and the other is an effective mass $M_{\rm eff}^{\rm Ext}$ associated with the exterior region $ r > R$, extending up to spatial infinity.

The effective mass contained in a region $ r < \delta$ can be calculated using Whittaker's theorem~\cite{whit,marsh}, according to which 
\begin{equation}\label{dseff}
M_{\rm eff}^{\rm{Int}} = \frac{1}{4\pi} \, \oint _{v_2} dv_2 \, V_{\, , i} \, n^i ~, \quad n^i = (V, 0,0) 
\end{equation}
where, in spherical polars,  $dv_2 = \delta^2 \, {\rm sin}\theta \, d\theta \, d\phi $ and $V^2$ denotes the temporal component of the metric tensor, as defined by the invariant line element $ds^2 =  V^2 dt^2 - g_{ij}dx^i dx^j $.

In the case of the $RN$ metric (in units of the gravitational constant $G=\frac{1}{8\pi}$)
\begin{equation}\label{rmmetric}
V^2 = 1 -\frac{2M}{r} + \frac{Q^2}{r^2}~.
\end{equation}
with $M$ the Schwarzschild mass (\ref{delMfinal}) and $Q$ the charge, 
one has
\begin{equation}\label{rn1}
V = \big(1 - \frac{2M}{r} + \frac{Q^2}{r^2} \big)^{1/2}~.
\end{equation}
On substituting in (\ref{dseff}) with $\delta=R$,  
one obtains~\cite{marsh}:
\begin{equation}\label{minner}
M_{\rm eff}^{\rm Int}= M - \frac{Q^2}{R}~.
\end{equation}

The effective mass associated with the exterior region $r > R$, has also been calculated in \cite{marsh} and the result is:
\begin{equation}\label{mext}
M_{\rm eff}^{\rm Ext} = \frac{Q^2}{R}~,
\end{equation}
so that the sum 
\begin{equation}\label{tot}
M_{\rm eff}^{\rm tot}  \equiv M_{\rm eff}^{\rm Int} + M_{\rm eff}^{\rm Ext} = M~.  
\end{equation}
The reader should notice the negative mass contribution in the right-hand-side of (\ref{minner}), which is compensated by the respective positive contribution in the effective mass of the exterior region (\ref{mext}), so that the sum $M_{\rm eff}^{\rm tot}$ (\ref{tot}) yields the Schwarzschild mass $M$ of the RN black hole.  The effective mass (\ref{minner}) appears~\cite{marsh} in the expression for the radial acceleration of a  neutral test particle falling into the RN black hole ($\tau$ is the proper time):
\begin{equation}\label{acc}
\frac{d^2\, r}{d\, \tau^2} = -\frac{1}{r^2}\big(M-\frac{Q^2}{r}\big)~,
\end{equation}
implying that the gravitational field, which in general varies with the distance $r$, becomes \emph{repulsive} when the effective mass $M-\frac{Q^2}{r}$ becomes negative at $r < Q^2/M$. Thus, neutral matter falling into the RN black hole will ultimately accumulate on a (2+1)-dimensional hypersurface for which the effective mass $M-\frac{Q^2}{r} =0$.

One may then attempt to apply the above considerations of \cite{marsh} on this surface, by defining appropriately the two types of effective mass discussed previously.
However, there are important differences. First, in view of the deficit angle $\eta^2$, the total energy in the exterior region is 
divergent formally and we cut such a divergence off using a bag model, and second, the inner region enclosed by the spherical surface of radius $\delta$ (\ref{delM}) is described by a different (de Sitter) space-time, 
which acts as a regulator of the RN singularity. Nonetheless, there are some features of our regularised solution that are in qualitative agreement with the RN analysis of \cite{marsh}
as far as the geometrical r\^ole of the charge is concerned. Let us see what implications such considerations have on our understanding of relations like (\ref{ratio}) that we have found above to characterise our solution.

In our problem, in the exterior region we do have a RN-like geometry (\ref{ab1}), with the (magnetic monopole) charge $Q^2$ being sourced by the KR axion charge: 
\begin{equation}\label{charge}
Q^2 = 2 \zeta^2 = g^2
\end{equation}
on account of (\ref{pdef}).  In the context of our bag-like model of the monopole~\cite{msmono}, the total energy (\ref{energy3}) when expressed in terms of $\delta$ (\ref{delMfinal}), using (\ref{charge}), yields
\begin{equation}\label{energydelta}
{\mathcal E}\simeq 1.13 \, \frac{Q^2}{\delta}~.
\end{equation}
The effective mass contained in the region $ r < \delta$ can be calculated using Whittaker's theorem~\cite{whit,marsh}(\ref{dseff}), using the de Sitter metric (\ref{distr}), 
for which 
\begin{equation}\label{ds2}
V = \big(1 - \frac{1}{3}\, \Lambda \, r^2 \big)^{1/2}~,
\end{equation}
with $\Lambda$ given in (\ref{Lident}). 
Then, from (\ref{dseff}) we obtain a \emph{negative} effective mass (consistent with the repulsive gravitational nature of de Sitter space time)
\begin{equation}\label{dseff2}
M_{\rm eff}^{\rm{Int}} = -\frac{1}{3}\, \Lambda\, \delta^3 < 0~.
\end{equation} 
If one requires (as in the RN case) that the sum of both exterior and interior effective masses should equal the Schwarzschild mass ((\ref{tot}))
then
in our case one should have
\begin{equation}\label{mext2}
M_{\rm eff}^{\rm Ext} = \frac{Q^2}{\delta}
\end{equation}
which is consistent with (\ref{mext}) for $R=\delta$. From (\ref{delMfinal}), (\ref{dseff2}) and (\ref{mext2}), one can verify the validity of
(\ref{tot}) since
\begin{equation}\label{variousM}
M_{\rm eff}^{\rm Ext} = \frac{3}{2} \, M \quad \rm{and}  \quad  M_{\rm eff}^{\rm Int} = -\frac{1}{2}\, M~.
\end{equation}

 From (\ref{energydelta}) and (\ref{mext2}) we observe that we obtain a value for the total energy ${\mathcal E}$ in the shell slightly larger than the effective mass
$M_{\rm eff}^{\rm Ext}$ in the region exterior to the core~\footnote{The reader should recall that  the total energy (\ref{energydelta}), in terms of the Schwarzschild mass (\ref{delMfinal}), is given by (\ref{ratio}) (which also stems from (\ref{energydelta}), (\ref{mext2}) and (\ref{variousM})). In our discussion of the weak equivalence principle in this section, we relate the shell energy $\mathcal E$ to the exterior effective mass $M_{\rm eff}^{\rm Ext}$, rather than  the Schwarzschild mass $M$, since the latter contains contributions ({\it cf.} (\ref{tot})) from both the (de Sitter) core and exterior regions of the self-gravitating KR monopole.}.
This can be attributed to the non-zero contributions of the KR axion $b(x)$ and electromagnetic fields to the energy functional $\mathcal E$ (\ref{energy3}), as well as the gravitational self-binding energy.
Generically, in \emph{nearly flat space-times}, as is the case in the exterior region of the core, one can define the total energy of an extended object as the integral of the temporal component of the stress tensor over a spatial volume $E_{\rm total} = \int dV \, T_{00}$. Under the 
assumptions that the components of the total momentum vector of the system are zero and the object is considered ``quasi-static'', that is, there is no significant energy present in the form of gravitational waves, then the object's  `mass' $\tilde m$ can be defined as (in units of the speed of light in vacuo): 
\begin{equation}\label{mtilde}
\tilde m = E_{\rm total} + E_{\rm binding}~,
\end{equation} 
where $E_{\rm binding} < 0$ denotes the  Newtonian gravitational self-binding energy. In our case, for large core radius, the criterion of (approximate) space-time flatness, along with the other assumptions, is satisfied; one may  thus identify $E_{\rm total} = {\mathcal E}$ and $\tilde m=M_{\rm eff}^{\rm Ext}$ in (\ref{mtilde}). Hence, the fact that the total energy is larger than the mass is naturally explained. 

We also note that, for us, the role of the infalling neutral matter is played by the KR axion pseudoscalar field, which thus will accumulate on the surface of radius $\delta$, since it is on this surface that the radial acceleration will vanish~(\ref{acc}).

We can summarise that, in view of the negative effective mass contributions of the de Sitter regulator, the weak equivalence principle, where one would equate
the total energy $\mathcal E$ with the inertial mass, fails. However, this should be expected for gravitating \emph{extended} objects, as is our case, given that the weak equivalence principle characterises point-like masses. Nevertheless, since the core radius is very large (compared to the Planck length), it is expected that such discrepancies will be small, since gravitational effects in the exterior of the core would be suppressed. This is indeed the situation that characterises our  case, where the gravitational total energy (\ref{energydelta}) is found to be almost the same magnitude as the effective mass of the monopole (\ref{mext2}). From the approximate validity of the weak equivalence principle, we can conclude that the motion of our monopole is akin to a point particle.  

It is interesting to note, though, that the weak-equivalence principle can be accommodated exactly through 
a choice of the value of our regularising cosmological constant $\Lambda$ in the core region, such that 
\begin{equation}\label{req}
\frac{\mathcal E}{M_{\rm eff}^{\rm Ext}} =1, \quad M_{\rm eff}^{\rm Ext} = \frac{Q^2}{\delta}~.
\end{equation}
In this scheme, the positive contributions to the total energy functional from the cosmological constant, KR axion, and electromagnetic fields will screen the negative binding energy due to the  gravitational effects (\emph{cf.} (\ref{mtilde})), leading to (\ref{req}).  This choice of regularisation scheme is consistent with a negative dilaton potential in the de Sitter region, where the string theory (the Ultra-Violet (UV) completion of our low energy model) is strongly coupled and such a potential might be generated, for example, through non-perturbative string-loop corrections. Outside this region, where string theory is weakly coupled and our low energy model is an effective description of the dynamics, the tree level dilaton potential \emph{vanishes} due to arguments based on conformal invariance~\cite{string}. In this way, our asymptotic solutions (which are valid outside the de Sitter region) are \emph{not} affected. Consequently we can introduce, instead of (\ref{Lident}),  a regularised $ \Lambda^\xi$: 
 \begin{equation}\label{Lident2}
 \Lambda^\xi = \frac{1}{4}\xi^2 \,  \,  \lambda \, \eta^4 > 0, \quad {\rm for} \quad  \lambda > 0~,
 \end{equation}
with $\xi \in {\tt R}$ a real number to be determined. We obtain from (\ref{delM}) and (\ref{param}):
\begin{eqnarray}\label{delM2}
\delta & \simeq &  2^{3/4} \, |\xi |\, \frac{|\zeta |}{\eta} \, \big(\zeta^2 \, \lambda \big)^{-1/4} ~, \nonumber \\
m^{\rm \xi}  & \simeq & 8\pi \, \Big(\frac{2^{1/4}}{|\xi|} - \frac{1}{6} 2^{5/4} \, |\xi|^5\Big) \, |\zeta| \, \big(\lambda \, \zeta^2 \big)^{1/4} \, \eta ~,
\end{eqnarray}
and from (\ref{dident}) we also obtain
\begin{equation}\label{alfa2}
0 < \alpha = 1.26 \, \xi^2 \, \big(\lambda \, \zeta^2 \big)^{-1/2}  \ll 1~,
\end{equation}
since we expect $|\xi| = {\mathcal O}(1)$, given that the role of $\xi$ is to set the ratio in (\ref{ratio}) to one. On account of (\ref{Lident2}), we obtain for the energy (\ref{energy2}):
\begin{equation}\label{energy4}
{\mathcal E}^{\rm \xi} \simeq  1.13 |\xi^2| \frac{Q^2}{\delta} 
\end{equation}
instead of (\ref{energydelta}). 
Hence, requiring 
\begin{equation}\label{xivalue}
\xi^2 = \frac{1}{1.13} \simeq 0.89
\end{equation}
we ensure (\ref{req}). This implies a smaller cosmological constant than previously chosen (\ref{Lident}).

\section{Conclusions}

In this paper we have regularised the curvature singularity characterising our recent self-gravitating  KR electromagnetic monopole solution~\cite{msmono}.
 The regularisation was achieved by cutting off the singular region by means of a de Sitter space implied by the scalar sector of the theory. The pertinent positive cosmological constant was proportional to the fourth power of the spontaneous symmetry breaking scale $\eta$ of the internal SO(3) symmetry of the model. We employed Israel-junction conditions when matching this interior region with the Reisssner-Norstr\"om (RN) black-hole space-time that characterises the outer region. This regularisation and our earlier bag-like structure~\cite{msmono}  allows, in a self-consistent way, for a calculation of both the finite total energy ${\mathcal E}$ of the monopole, and the mass coefficient $m$ appearing in the appropriate gravitational potential term of the RN metric. Notably in this case, $m$ cannot be identified necessarily with the total energy, due to the divergent infrared behaviour of the latter. This feature also characterises the global monopole case~\cite{vilenkin}, in the absence of the KR charge. However, in this latter case, the above matching leads to a negative gravitational mass coefficient  $m < 0$ and thus repulsive gravitational effects~\cite{lousto}.
 
 By contrast, in  our KR electromagnetic monopole case, the gravitational mass coefficient $m$ turns out to be positive, but proportional to the KR axion charge, whose role is thus crucial in ensuring normal (attractive) gravitational effects of the KR monopole. The pertinent calculations of the total energy ${\mathcal E}$ of the monopole  and $m$ have been performed within a phenomenologically interesting regime of the parameters of the model, in which the scale $\eta$ is assumed much smaller than the Planck scale, since our interest is to consider monopoles with masses within the range of current or future colliders.  When employing a regularisation using a  de Sitter space-time region  with the value of the cosmological constant used in the global monopole case, we find that ${\mathcal E} \ne m$.(Although both quantities in the ratio are of the same order of magnitude, with values very close to each other.)
This small violation of the weak equivalence principle is to be expected since we deal in our case with gravitational effects that have non-trivial contributions to the effective mass. 
We put our result into context by comparing with studies of conventional RN solutions~\cite{marsh}; we can obtain agreement 
between the two approaches, in the sense of (\ref{tot}), by choosing a cosmological constant which is slightly reduced in this case as compared to that of the global monopole case~\cite{lousto}. We have speculated that, from a microscopic view point, such a regularisation might arise from a non-trivial dilaton dynamics inside the core de Sitter region, giving rise to a negative dilaton potential in that region. Outside the region, the dilaton is stabilised to a  constant value, 
corresponding to a zero value of its potential, and thus the asymptotic solutions of ref.~\cite{msmono} are not affected.

 We reiterate that, the interpretation of the de Sitter space regularisation of our monopole core as a repulsive gravitational force, allows us to understand the stability of the self-gravitating KR solution; the stability is  a consequence of a balance of these repulsive forces with the attractive gravitational forces of the positive mass parameter of the RN space-time in the exterior region, at the radius $\delta$ where the Israel matching conditions of the two metrics are enforced. The non-trivial KR-axion charge $\zeta$ of our model is crucial to produce this balance. 

By considering, in detail, phenomenologically realistic microscopic scenarios for the KR monopole it would be possible to determine the symmetry breaking scale $\eta$ from first principles, and thus make definite predictions for the monopole mass. Moreover, the role of the various fields, like the scalars $\chi^A$, would also be elucidated, and this will lead to a better understanding of the production mechanism of such monopoles. Since the KR electromagnetic monopoles are composite, non-point-like, objects, their production at colliders is expected to be strongly suppressed, according to generic arguments ~\cite{drukier}. However, non-suppressed production may be expected for such objects in environments with high temperature and/or strong external magnetic fields, as a consequence of a thermal Schwinger-like mechanism, according to recent arguments~\cite{gould}. Hence, the demonstration from first principles of how a low mass KR monopole arises in the physical spectrum of microscopic models, is a pressing issue, not only of theoretical, but also of direct experimental relevance. 
We hope to attempt to answer (some of) the above questions in the future.

\section*{Acknowledgements}

The work of N.E.M. and S.S. is partially supported by STFC (UK) under the research grant ST/P000258/1.

\end{document}